\documentclass[conference]{IEEEtran}
\IEEEoverridecommandlockouts
\usepackage{cite}
\usepackage{amsmath,amssymb,amsfonts}
\usepackage{algorithmic}
\usepackage{graphicx}
\usepackage{textcomp}
\usepackage{xcolor}
\begin{document}

\title{A Brief History of Space VLBI\\
}

\author{\IEEEauthorblockN{Leonid I. Gurvits}
\textit{Joint Institute for VLBI ERIC (JIVE)} \\
Dwingeloo, The Netherlands \\
and \\
\textit{Faculty of Aerospace Engineering} \\
\textit{Delft University of Technology}\\
Delft, The Netherlands \\
ORCID: 0000-0002-0694-2459
}

\maketitle

\begin{abstract}
Space Very Long Baseline Interferometry is a radio astronomy technique distinguished by a record-high angular resolution reaching single-digit microseconds of arc. The paper provides a brief account of the history of developments of this technique over the period 1960s–2020s.
\end{abstract}


\section{Introduction}

The year 2023 marks the 90th birthday of radio astronomy: it is commonly accepted that this large branch of astronomy had been ``launched'' with a column on the front page of New York Times of 5 May 1933 describing the discovery of ``cosmic noise'' by Karl Jansky \cite{Jansky-1933N}. For almost two thirds of this period, since the mid-1960s, a radio astronomy technique called Very Long Baseline Interferometry (VLBI) firmly holds the record in angular resolution when viewing celestial objects (provided they emit in the radio domain of the electromagnetic spectrum). The first experimental demonstration of this technique took place in 1967 by three US and one Canadian groups (see \cite{TMS-2017}, section 1.3.14, and references therein). The technique was discussed two years earlier in \cite{MKSh-1965}. Interestingly, the latter publication in its draft version of 1963 contained a paragraph referring to a potential placement of a radio interferometer antenna on a spacecraft aiming to achieve an angular resolution fundamentally impossible for Earth-based instruments. This paragraph was removed from the final version at the request of strict censorship in the USSR of all topics dealing with space exploration at that time. The latter was independently confirmed to the author in four private communications by Lev~A.~Lebedev (1987), Gennady~Sholomitsky (1991), Nikolai~Kardashev (2016) and Leonid~Matveenko (2018). So, if one takes into account this first mention of Space VLBI in the first half of 1960s, by now, the topic does have a respectable history. 

In the case of reflecting antennas (in a widely used professional slang -- ``dishes''), the resolution is defined by the diffraction limit, $\lambda/D$, where $\lambda$ is a wavelength, and $D$ -- a diameter of the reflector, just as is the case in the ``traditional'' optical astronomy. For typical in radio domain decimeter to meter wavelengths, affordable dishes of tens of meters in diameter can reach the angular resolution of tens of minutes of arc, far worse than a typical angular resolution of Earth-based optical telescopes, which is of the order of a second of arc. A concept of interferometers was introduced in radio astronomy in the beginning of the 1950s \cite[chapter~5]{BFB+FGS-2014}. It enabled radio astronomers to sharpen the angular resolution defined by the same simple formula mentioned above, with a substitution of the antenna diameter with the projection on the image plane of the baseline connecting two elements of the interferometer. Later, in the second half of the 1960s, the concept of radio interferometry got its ultimate extension to the baselines comparable to the Earth diameter. This led to the introduction of VLBI mentioned above. VLBI offers a breathtaking sharpness of viewing reaching microarcseconds. However, very soon radio astronomers realised that this technique has its limits: there are celestial radio sources that remained point-like (unresolved) even in observations with the global VLBI arrays the size of our planet.  As is clear from the simple formula at the beginning of this paragraph, the only solution of getting an even sharper resolution at a given wavelength than achievable with the baseline comparable to the Earth diameter is to place at least one element of an interferometer in space, i.e, creating a Space VLBI (SVLBI) system.

\section{Prehistory of SVLBI}

Once the idea of radio interferometers larger than the planet had been pronounced, several project studies took off in Europe, USA and USSR in the 1970s (see a brief review of the state of affairs in these activities as of 1984 in \cite{Burke-1984IAUS}). Among various engineering components of these projects, the emphasis was placed on the antennas of spaceborne VLBI elements. Some of these studies were rather futuristic, like the ``infinitely expandable speceborne radio telescope'' \cite{Buyakas+1979} which, in modern terms, can be called a square kilometre telescope in space. At the other end of the antenna-size scale was a proposal to use a $\sim$3-m communication antenna of the proposed NASA Venusian mission \textit{VOIR} as a VLBI antenna during the mission's cruise phase, thus operating with baselines ranging from several to $\sim$100 million km \cite{Burke-1982OVLBI}.

Several early SVLBI studies proceeded to advanced stages and played important roles in providing inputs into later projects that resulted in operational SVLBI missions. Arguably, among these advanced studies the most prominent were the following three:
\begin{enumerate}
\item \textit{Quasat}, a joint ESA--NASA study of an orbiting $\sim$15-m radio telescope for imaging SVLBI observations at dm–cm wavelengths with angular resolution sharper by a factor of 3–5 compared to global Earth-based systems \cite{ESASP213}. The project was under study in the first half of the 1980s and reached the ESA Phase A study \cite{ESASCI88-4}.
\item \textit{RACSAS} (Radio Astronomical Space System for Aperture Synthesis), a Soviet project of a low-orbit VLBI synthesis system with a 30-m spaceborne radio telescope \textit{KRT-30}, which was under development from the late 1970s through the 1980s \cite[section 7.3.3]{Davies-1997}. The \textit{KRT-30} telescope was designed as a multi-task facility for remote sensing and radio astronomy for potential deployment in space by the Soviet spaceplane \textit{Buran} \cite[p.~383]{Hendrickx+Vis-2007}, an analogue of the US \textit{Space Shuttle}.
\item \textit{IVS} (International VLBI Satellite), a joint project study by ESA, NASA and Soviet Space Agency in the second half of the 1980s \cite{IVS-1991}. It was an ambitious project of a 25-m speceborne telescope capable to observe as a VLBI element at wavelengths from 6~cm down to 1.5~mm from a range of orbits with a varying apogee from 20,000~km to 150,000~km. 
\end{enumerate}
None of the above projects has materialised. But many components of design studies of hard- and soft-ware of these projects found their way into real SVLBI systems which began operations in the middle of the 1980s. The early studies also served as a ``training ground'' for the human workforce of the implemented SVLBI missions in the following decades.

\section{Space VLBI becomes a reality}

\subsection{First SVLBI ``fringes''}
\begin{figure}[htbp]
\centerline[{\includegraphics[width=0.48\textwidth]{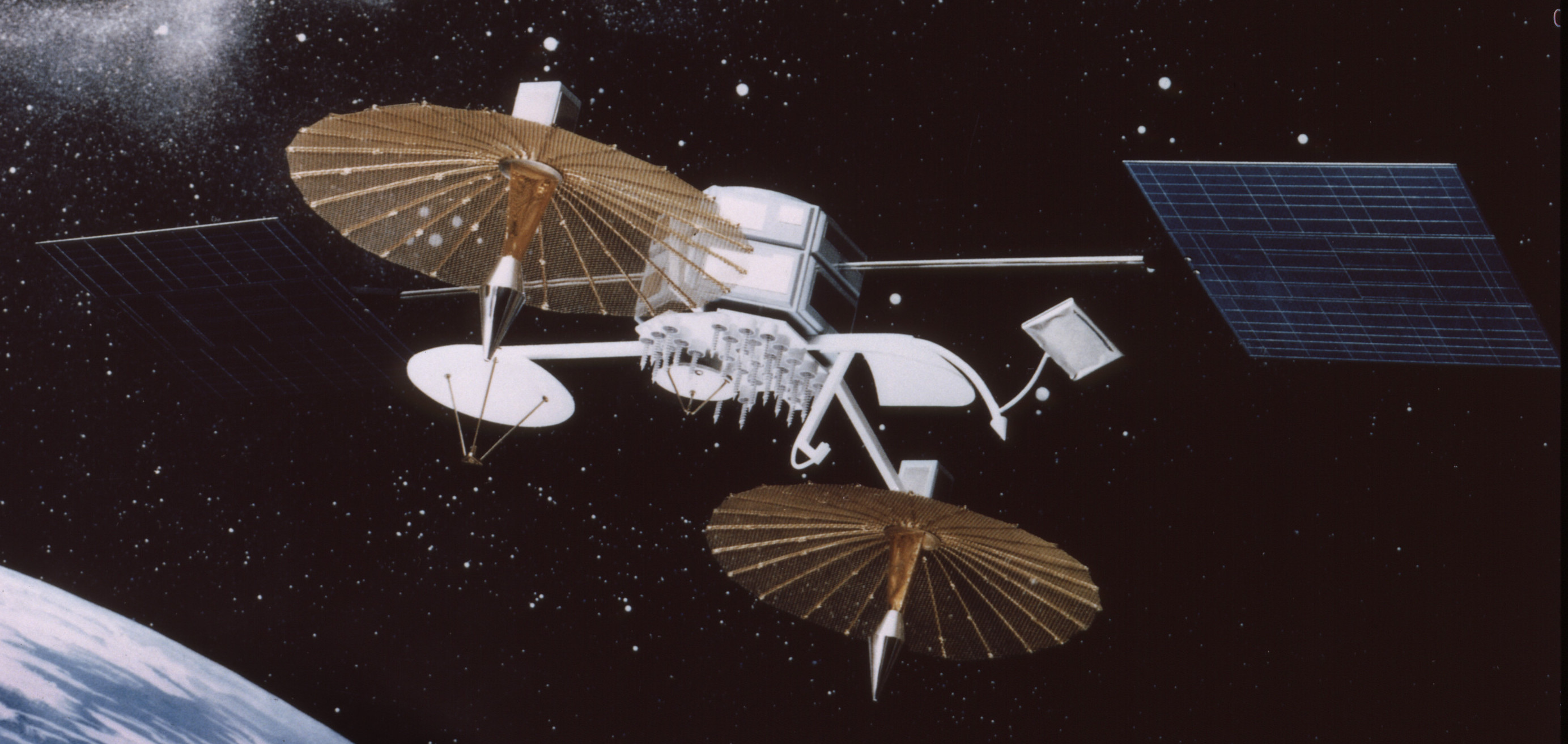}}
\caption{Tracking and Data Relay Satellite of the 1st generation, an artist's impression. \textit{Credit}: NASA.}
\label{f:TDRS}
\end{figure}

The first ad hoc experimental demonstration of SVLBI was conducted with the NASA’s geostationary Tracking and Data Relay Satellite System (\textit{TDRSS}) in 1986 \cite{Levy+1986Sci}, Fig.~\ref{f:TDRS}. Its geostationary satellite was equipped with two 4.9~m antennas, one of which was used in this demonstration as a radio telescope. The observations were conducted at 13~cm and 2~cm together with a network of large Earth-based antennas in Australia, Japan and the US. 

The space-borne component of this first SVLBI system was not designed as a radio telescope, and was not supposed to operate as such. It was the ingenuity of the NASA engineers that made possible re-configuring and re-programming the on-board instrumentation making it compatible with the Earth-based VLBI radio telescope and data processing systems. 

The main achievement of the \textit{TDRSS} Orbital VLBI experiment was an experimental confirmation of the feasibility of obtaining an interferometric response (so called ``interferometric fringes'') on baselines exceeding the Earth diameter and involving a radio telescope not fixed on the surface of Earth. The maximum projected baseline achieved in the demonstration was $\sim$2.2~Earth diameters (ED). It has to be noted that the sensitivity of an interferometer to a source’s brightness, traditionally measured in radio astronomy in degrees Kelvin (of a hypothetic black body's temperature of the same brightness as the source) depends on the physical length of the projected baseline. In other words, the source can be resolved with a given baseline only up to a certain brightness temperature. If the brightness temperature is higher for a given total flux density of the source, a longer baseline is needed to resolve the source’s image. The \textit{TDRSS} Orbital VLBI experiment indicated that in several quasars the brightness temperature is higher than can be measured with Earth-based VLBI arrays.

\subsection{Muses-B, VSOP, HALCA: the first dedicated operational SVLBI facility}
In the wake of the success of the \textit{TDRSS} VLBI demonstration experiment, the VLBI Space Observatory Program (VSOP) was conceived at the Institute of Space and Astronautical Science in Japan in the end of the 1980s \cite{Hirax+1998Sci}. Although the project was formally aimed primarily at testing the new four-stage solid fuel rocket \textit{M-V} and new satellite platform, christened \textit{HALCA} (Highly Advanced Laboratory for Communication and Astronomy) after the launch on 12 February 1997, it did have a core science program focused on SVLBI experiments.

\begin{figure}[htbp]
\centerline[{\includegraphics[width=0.48\textwidth]{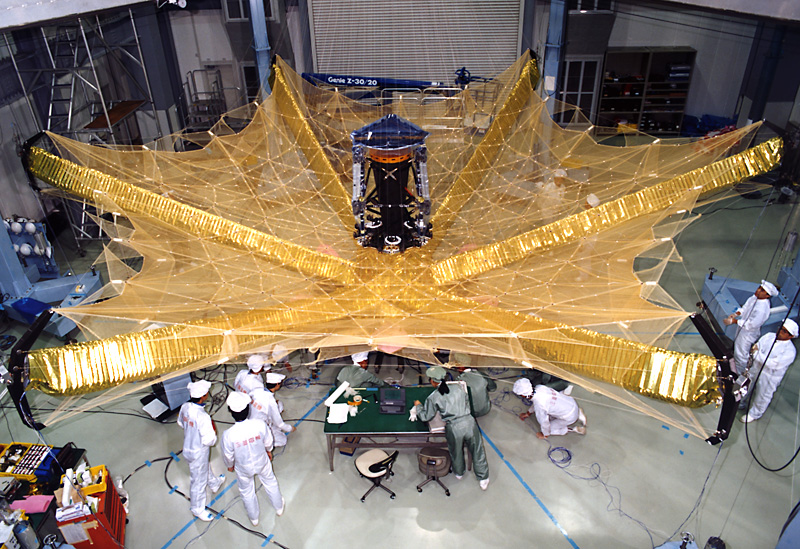}}
\caption{The \textit{HALCA} antenna deployment test at the Institute of Space and Astronautical Sciences. \textit{Credit}: ISAS/JAXA.}
\label{f:HALCA}
\end{figure}

The main component of the mission was the space-borne radio telescope with a parabolic reflector of 8.8~m effective diameter, Fig.~\ref{f:HALCA}. Its surface was formed by a gold-coated molybdenum wire mesh. Three receivers enabled observations at 18, 6 and 1.3~cm in left-hand circular polarisation (LCP). The satellite was on a medium eccentricity orbit with the apogee of about 22,000~km above Earth and orbital period of about 6~hrs. The mission had a sizeable involvement of international community coordinated by a group called VISC (\textit{VSOP} International Science Council). Among other things, VISC coordinated such mission implementation issues as operation of five dedicated Earth-based tracking and data acquisition stations (one each in Australia, Japan and Spain, plus two stations in the US), orbit determination, VLBI data processing, Earth-based VLBI network arrangements and the policy for access to the mission by the worldwide science community. 

\textit{VSOP} operated with the aggregate data rate of 128~Mbit/s -- the highest sustainable data downlink rate of a space science mission at the time. The signal coherency -- the major requirement of a VLBI system -- was supported by a phase-locked loop (PLL), fed by a hydrogen maser frequency standard at a tracking station. Two of the three observing bands, 18 and 6~cm, operated successfully. The signal in the third, the highest frequency band of 1.35~cm, had a very high system noise, preventing science operations. It is believed that the reason was damage to the waveguide of this band due to a high level of vibration during the satellite launch. Three data processing (correlation) facilities supported the mission. They were located at the Dominion Radio Astrophysical Observatory in Penticton, Canada; National Astronomical Observatory of Japan in Mitaka, Japan; and National Radio Astronomy Observatory in Socorro, NM, USA. The \textit{HALCA} satellite operated for six years until 2003 and was switched off after its attitude control reaction wheels stopped operating after exceeding their warranty
 period significantly.

The \textit{VSOP HALCA} mission provided a wealth of observational material on continuum (both 18~cm and 6~cm wavelengths) and hydroxyl spectral line observations at 18~cm on compact structures of galactic and extragalactic radio sources at the hitherto unchartered sub-milliarcsecond angular scale. The mission was highly acclaimed: it was awarded with the Team Achievement Award of the International Academy of Astronautics in 2005. 

\vspace{-5pt}
\subsection{RadioAstron}
The next and so far the only other first-generation SVLBI mission is the Russian-led \textit{RadioAstron} \cite{NSK+2013ARep}. The project was conceived in the late 1970s at the Space Research Institute of the USSR Academy of Sciences and was launched from Baikonur on a \textit{Zenit-2SB} rocket and a \textit{Fregat} booster stage on 18 July 2011. The Astro Space Center of the Lebedev Physical Institute of the Russian Academy of Sciences and the Lavochkin Science and Production Association led the mission. 

\begin{figure}[htbp]
\centerline[{\includegraphics[width=0.48\textwidth]{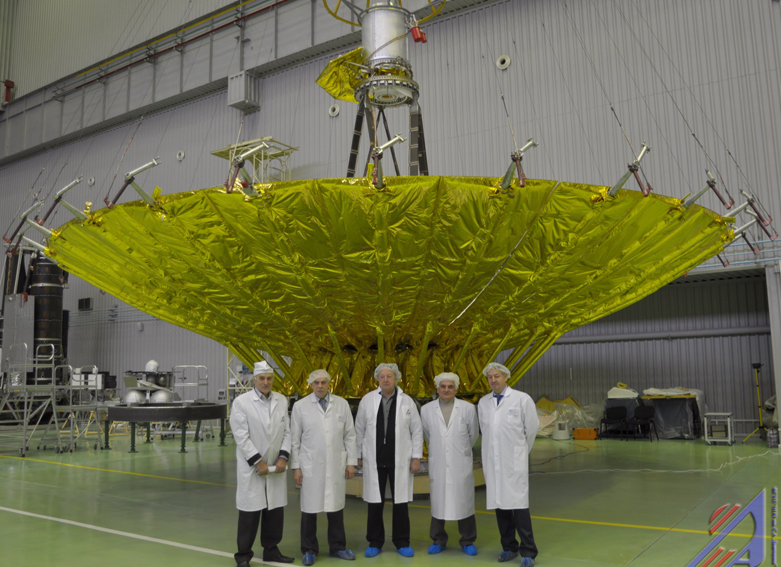}}
\caption{The \textit{RadioAstron Spektr-R} 10~m antenna at the Lvochkin Association assembly facility. In front of the satellite -- a group of leading engineers of the \textit{Spektr-R} project. \textit{Credit}: Lavochkin Science and Productions Association.}
\label{f:Spektr-R}
\end{figure}

The main component of the RadioAstron’s science payload was the 10-m parabolic reflector formed by 27 carbon-fibre petals and a central circular part, Fig.~\ref{f:Spektr-R}. The telescope was equipped with four radio astronomy dual-polarisation (both left- and right-hand, LCP and RCP) for the wavelengths 92, 18, 6 and 1.3~cm. The satellite operated on a highly eccentric evolving orbit with its apogee reaching the distance of 350,000 km from Earth ($\sim$28~ED) and with an orbital period of about 9~days. Just as \textit{VSOP} did, \textit{RadioAstron} was able to use a highly stable heterodyne signal of an Earth-based hydrogen maser frequency standard through a PLL. However, the \textit{RadioAstron} spacecraft also carried two active Hydrogen maser standards. One of these devices was used as the prime provider of the highly stable heterodyne signal for 6 years, twice longer than the warranty period, from the launch until July 2017. Just as in the case of \textit{VSOP}, \textit{RadioAstron} had a large involvement of international community in its mission operational and science components. Similarly to \textit{VSOP}, the \textit{RadioAstron} International Science Council (RISC) coordinated the global international involvement in the mission. The Earth-based support was provided by two data acquisition stations in Pushchino (Russia) and Green Bank (WV, USA). They received the raw VLBI data stream from the satellite with the data rate of 128~Mbit/s in the same data format as \textit{VSOP}. The \textit{RadioAstron} VLBI data were processed at three correlators operated at the Astro Space Center in Moscow (Russia), Max-Plank-Institut für Radioastronomie in Bonn (Germany) and Joint Institute for VLBI ERIC in Dwingeloo (The Netherlands).

After completion of the in-orbit checkout and verification, the mission has become open to the worldwide science community under the ``open sky'' access policy. The \textit{RadioAstron} mission has issued 6 Calls for Science Proposals, each covering one operational year. Due to the achievable baselines reaching 28\,ED, the \textit{RadioAstron} mission has established angular resolution records in all four operational bands, and the absolute record of about 7 microarcseconds at the shortest observing wavelength of 1.3~cm. The mission was completed in 2019, after operating in orbit nearly 7.5~years, 2.5 times longer than the industrial warranty of the spacecraft.

\section{SVLBI Legacy}

The two first-generation SVLBI missions, \textit{VSOP} and \textit{RadioAstron}, which followed the first demonstration SVLBI experiment with \textit{TDRSS}, created a solid technological basis for further development of space-borne radio interferometry. Table \ref{t:SVLBI-1} summarises their major technical parameters.

\vspace{-13pt}
\begin{table}[htbp]
\caption{Major specifications of the implemented to date SVLBI systems}
\begin{center}
\begin{tabular}{|l|c|c|c|}
\hline
\textbf{ } & \textbf{\textit{TDRSS}}& \textbf{\textit{VSOP}}& \textbf{\textit{RadioAstron}} \\
\hline
In-orbit operations & 1986--1988 & 1997--2003 & 2011--2019 \\
\hline
Diameter of antenna [m] & 5.8.         & 8.8          & 10               \\
\hline
$B_{\rm max}$ [ED]$^{\mathrm{a}}$ & 2.2   &  3   & 28                   \\
\hline
Wavelengths [cm]  & 13, 2 & 18, 6 & 92, 18, 6, 1.3 \\
\hline
Data rate [Mbps]    & 28  & 128 & 128   \\
\hline
\multicolumn{4}{l}{$^{\mathrm{a}}$Maximal baseline projection on the image plane} \\
\multicolumn{4}{l}{\,\,\,in units of Earth Diameters [ED].}
\end{tabular}
\label{t:SVLBI-1}
\end{center}
\end{table}
\vspace{-10pt}

In a broad sense, the radio astronomy facilities operating at the highest angular resolution address three major science topics.
\begin{enumerate}
\item Physics of compact continuum sources, primarily extragalactic objects, associated with Active Galactic Nuclei (AGN) powered by supermassive black holes. The existing paradigm of AGN considers the synchrotron mechanism as the main origin of their electromagnetic emission. However, important details of these extremely powerful emitters remain enigmatic. The key to understanding the physics of these ``most powerful engines'' of the Universe lies in their concentration in compact areas, parsecs and sub-parsecs in linear size. To zoom into these objects, located at the distances of mega- and gigaparsecs from the Earth, one needs to achieve the angular resolution going down to microarcseconds. 

\item Ultra-compact sources of narrow-band spectral line emission of cosmic masers. At present, most ``popular'' molecules responsible for maser emission are water vapour H$_2$0, emitting at 1.35~cm, and hydroxyl OH, emitting at 18~cm. For these molecules and sources, both galactic and extragalactic, the Earth-based VLBI studies are limited in resolution, insufficient for detailed zoom into the areas of maser emission generation.

\item Radio signal propagation through intergalactic, interstellar, interplanetary, and near-Earth media. These media represent an important constituency of cosmos at various scales, from cosmological to planetary scales.
\end{enumerate}

The science legacy of the \textit{VSOP} mission is presented in details in \cite{HEM-2000proc} and \cite{HFTM-2009ASPC}. While data processing and evaluation of scientific results of the \textit{RadioAstron} mission are still ongoing at the time of this writing, some results are presented in \cite[and references therein]{AdSpR-2020}. A brief set of lessons learned so far on the development end exploitation of SVLBI missions is presented in \cite{LIG-2020AdSpR}. The science case of ultra-high angular resolution in radio astronomy is enhanced by the two dedicated SVLBI missions implemented to date. Together with recent developments in Earth-based radio astronomy, they justify the inevitability of future SVLBI facilities in all domains of the radio spectrum, from ultra-long (tens of meters and longer) to sub-millimeter wavelengths \cite[and references therein]{AdSpR-2020}. 

\section*{Acknowledgments}
The author is grateful to Bob Campbell (JIVE) for very useful comments. The TDRS OVLBI demonstration resulted from the efforts of the TDRSS organisation and cooperating observatories in Australia, Japan and USA. The VSOP Project was led by the Institute of Space and Astronautical Science (Japan) in cooperation with many agencies, institutes, and observatories around the world. The RadioAstron project, a collaboration of partners in Russia and other countries was led by the Astro Space Center (Lebedev Physical Institute) of the Russian Academy of Sciences and the Lavochkin Association of the Roscosmos State Corporation.


\bibliography{Brief-history-SVLBIbib}   

\end{document}